\documentclass[twocolumn,
showpacs,preprintnumbers,amsmath,amssymb,floatfix,aps]{revtex4}

\usepackage[dvips]{graphicx}
\usepackage{dcolumn}
\usepackage{bm}

\begin{document}


\title{New asymptotic behaviour of the  surface-atom  force out
of thermal equilibrium}

\author{Mauro Antezza$^{a}$, Lev P. Pitaevskii$^{a,b}$ and Sandro
Stringari$^{a,c}$}

\affiliation{
$^{a}$Dipartimento di Fisica, Universit\`a di Trento
and BEC Center, INFM-CNR, Via Sommarive 14, I-38050 Povo, Trento, Italy \\
$^{b}$ Kapitza Institute for Physical Problems, ul. Kosygina 2, 119334 Moskow,
Russia\\
$^{c}$ \'Ecole Normale Sup\'erieure and Coll\`ege de France,\\
Laboratoire Kastler Brossel, 24 Rue Lhomond 75231 Paris, France}

\date{\today}

\begin{abstract}
The Casimir-Polder-Lifshitz force felt by an atom near the surface of a
substrate is calculated out of thermal   equilibrium in terms of
the dielectric function of the material and of the atomic polarizability.
The new force   decays like $1/z^3$ at large distances (i.e. slower than at equilibrium),
exhibits a sizable temperature dependence and is attractive or
repulsive depending on whether the temperature of the substrate
is higher or smaller than  the one of the environment. Our predictions can be relevant for
experiments with ultracold atomic gases. Both
dielectric and metal substrates are considered.
\end{abstract}
\pacs{34.50.Dy, 12.20.-m, 42.50.Vk, 42.50.Nn}

\maketitle

An intense effort has been devoted in the recent years
to study the force felt by an atom near
the surface of a substrate (see, for example, \cite{Babb,articolo1}
and references therein). These  studies are motivated both
by the possibility of technological applications \cite{capasso}
as well as by the fundamental search for stronger constraints on
hypothetical non Newtonian forces \cite{dimopulos}.
Experimental and theoretical   research has also recently  focused
on the forces acting on ultracold atomic gases,
including atomic beams \cite{Hinds,Aspect,Shimizu}, Bose-Einstein condensates
\cite{Vuletic,Ketterle,eric05,articolo1}
and degenerate Fermi gases \cite{Iacopo}.

The force generated by the surface contains in general two independent
components. The first one is related to zero-point fluctuations of the
electromagnetic field. At short distances $z$
(typically less than fractions of
microns) this force behaves like $1/z^{4}$ and is the analog of the van der
Waals-London interatomic force. At larger distances the inclusion of
relativistic retardation effects gives rise to a different $1/z^{5}$
dependence characterizing the so-called Casimir-Polder regime
 \cite{CP,LP,books}. Notice that both the van der Waals-London and
Casimir-Polder forces depend on the
temperature only through the dielectric
properties of the substrate. The resulting temperature
dependence is usually negligible.

The second component of the force is due to the  thermal fluctuations of
the electromagnetic field. This effect was   first considered by
Lifshitz \cite{lifshitzDAN} who employed the theory of electromagnetic fluctuations developed by Rytov  \cite{Rytov}. We will refer to it as to the  Lifshitz force.
At distances larger than the thermal photon wave length $\lambda _{T}=\hbar
c/k_{B}T$ (corresponding to $\sim 7.6\; \; \mu m$ at room temperature) this
force decays, at thermal equilibrium, like $1/z^{4}$, is
attractive, is proportional to the temperature and provides the leading
contribution to the total force. Conversely, the zero-point fluctuation
component prevails at smaller distances. In the present paper we are interested
in the thermal component of the force.
The Lifshitz force was originally evaluated  at full thermodynamic
equilibrium. A non trivial issue is the study of the force out of thermal
equilibrium \cite{Henkel1,cohen}, characterizing configurations
where the temperature of the
substrate and  the one of the surrounding walls located at large distances
(hereafter called environment temperature) do not coincide. The problem is relevant both for practical reasons, due to the possibility
of tuning the two temperatures independently, and
for a better understanding of the interplay between zero-point and thermal
fluctuation effects. A first important investigation of the
surface-atom force out of thermal equilibrium was
carried out by Henkel {\it et al.} \cite{Henkel1}
who calculated  the force generated by a dielectric
substrate at finite temperature by  assuming that
the environment  temperature  is zero.
The principal motivation of that paper was the study
of the force at short distances. 
In this Letter we show that, out of thermal equilibrium,
the force acting on the atom exhibits a new asymptotic behaviour,
characterized by a $1/z^3$ decay at large distances.

Let us   consider  an atom placed in vacuum at distance $z$
from the flat surface of a substrate made of a material with  dielectric function
$\varepsilon(\omega)=\varepsilon'(\omega)+i\varepsilon''(\omega)$.
We  choose a  coordinate system with the $xy$ plane coinciding
with the interface and the $z$ axis such that the substrate
occupies the region with $z<0$ and the vacuum the region with $z>0$.
In this Letter we assume that the substrate is locally at thermal
equilibrium at a temperature $T_S$ which can differ from the environment
temperature $T_E$, the global system being out
of thermal equilibrium, but in a stationary regime.
The total electromagnetic field will be in general the sum of the
radiation produced by the substrate and the one of the environment. In particular this latter radiation will be
partially absorbed and reflected from the substrate.
The forces produced by the $T_E$ and $T_S$ components
of the radiation add incoherently.
In typical experiments with ultracold atomic gases the environment temperature is determined by the chamber containing
the substrate and the trapped  atoms. We further treat the atoms as being at zero temperature
in the sense  that the surrounding radiation is not able
to populate their excited states which are assumed to be located at energies
$\hbar \omega_{at}$ much higher than the  thermal energy:
\begin{equation}
k_BT_S, \; k_BT_E << \hbar \omega_{at}\,.
\label{condition}
\end{equation}
This condition is very well satisfied at ordinary temperatures. Actually the first optical resonance of Rb atoms corresponds to $1.8 \; 10^4\;K$
In general the force  acting on a neutral atom can be written as
\cite{Gordon,Henkel1}
\begin{gather}
{\bf F}({\bf r})
\approx
\left\langle d_i^{\textrm{ind}}
\nabla E_i^{\textrm{f}}({\bf r})\right\rangle
+\left\langle d_i^{\textrm{f}}\nabla E_i^{\textrm{ind}}
({\bf r})\right\rangle
\label{forcepert}
\end{gather}
holding in lowest order  perturbation theory with $i=x,y,z$.
Here and in the following we use Einstein's summation convention.
The first term describes the  field fluctuations  correlated
with the induced atomic dipole moment while the second ones involves
the dipole fluctuations  correlated with the induced electric field.
At thermal equilibrium, where $T_S=T_E\equiv T$, the force can be
conveniently written in the form
\begin{equation}
F^{\textrm{eq}}(T,z)=F_{\textrm{0}}(z)+F_{\textrm{th}}^{\textrm{eq}}(T,z)
\label{Feq}
\end{equation}
where we have separated the contribution  $F_{\textrm{0}}(z)$
arising from the $T=0$  zero-point fluctuations    and the one
arising from the thermal fluctuations. At large
distances the $T=0$ force exhibits the Casimir-Polder asymptotic behaviour
\begin{equation}
F_{\textrm{0}}(z)_{z\to \infty} =
-\frac{3}{2}\frac{\hbar c \alpha_0}{\pi z^5}
\frac{\varepsilon_0-1}{\varepsilon_0+1}\phi(\varepsilon_0)
\label{Feqvacasymp}
\end{equation}
with the function $\phi(\varepsilon_0)$ defined, for example, in \cite{articolo1}. Conversely,
the thermal force approaches the Lifshitz law
\begin{equation}
F_{\textrm{th}}^{\textrm{eq}}(T,z)_{z\to \infty}  =
-\frac{3}{4}\frac{k_BT \alpha_0}{z^4}\frac{\varepsilon_0-1}{\varepsilon_0+1}
\label{neweqasymp}
\end{equation}
which then provides the leading contribution to the total force.
The asymptotic law
(\ref{neweqasymp}) is reached at distances larger
than the thermal wavelength $\lambda_T$.
In the above equations $\alpha_0$ ($=47.3 \times 10^{-24} cm^3$ for Rb atoms) and $\varepsilon_0$ are,
respectively, the static polarizability of the atom and the
static dielectric function of the substrate.
It is worth noticing that only the static optical
properties enter the  asymptotic laws (\ref{Feqvacasymp})
and (\ref{neweqasymp}), the corresponding dynamic
effects becoming important only at shorter distances.
It is also worth noticing that the  asymptotic behaviour
of the thermal force has a classical nature,
being independent of the Planck constant. The explicit behaviour of the force 
(\ref{Feq}) at all distances has been recently investigated   in  \cite{articolo1}.

Let us now discuss the behaviour of the force when
the system is not in equilibrium (neq).
Also in this case the force can be written
as the sum of the zero temperature contribution
$F_{\textrm{0}}(z)$ and of a thermal contribution which, however,
will differ from
$F_{\textrm{th}}^{\textrm{eq}}$ if $T_S\neq T_E$:
\begin{equation}
F^{\textrm{neq}}(T_S,T_E,z)=F_{\textrm{0}}(z)+
F_{\textrm{th}}^{\textrm{neq}}(T_S,T_E,z) \; .
\label{Fneq}
\end{equation}

The purpose of the present Letter is to calculate the force (\ref{Fneq})  and in particular  to exploit its behaviour at large distances.  Let us first consider the case of a   substrate  at finite temperature ($T_S=T\ne 0$) in the absence of the environment radiation ($T_E=0$). This problem was solved  by Henkel \emph{et al.} \cite{Henkel1} who obtained the  result
\begin{gather}
F_{\textrm{th}}^{\textrm{neq,ff}}(T,0,z)=
\frac{\hbar}{2\pi^2}\int_0^{\infty}\textrm{d}
\omega\;\frac{\varepsilon''(\omega)}{e^{\hbar\omega/k_BT}-1}\notag\\
\textrm{Re}\left[\alpha(\omega) \int_{V_S} \;
G_{ik}[\omega;{\bf r},{\bf r}_1]\;\partial_{z}G_{ik}^*[\omega;
{\bf r},{\bf r}_1]\;\textrm{d}^3{\bf r}_1\right]
\label{HenkelForce}
\end{gather}
for the thermal contribution to the force
originating from the fluctuations of the field (ff).
In eq.(\ref{HenkelForce})  $G_{ik}$
is the  Green function relative to the electromagnetic field.
The variable ${\bf r}_1$   should be integrated on the
volume $V_S$ occupied by the substrate which provides
the source of the thermal radiation.
The argument ${\bf r}$ instead defines the position
of the atom outside the substrate.
The Green function $G_{ik}$ then reduces  to its
transmitted component \cite{Sipe}.

The  force (\ref{HenkelForce}) contains
a repulsive coordinate independent \emph{wind} part (arising from the transmitted propagating modes)
proportional to $\alpha''(\omega)$ and produced by the
absorption of photons by the atom, and  a
\emph{dispersive} part (arising from the transmitted evanescent modes) proportional to $\alpha'(\omega)$.
Due to the condition (\ref{condition}) the \emph{wind} contribution
 can be
ignored and   the real part $\alpha'(\omega)$
can be replaced with its  static ($\omega=0$) value $\alpha_0$.
Furthermore also the
terms arising from the dipole fluctuations can be
ignored in the evaluation of the thermal force which can be conveniently rewritten in the form
  $F^{\textrm{neq}}_{\textrm{th}}(T_S,0,z)=4\pi\alpha_0\partial_z U_E(T_E,0,z)$ where $U_E=\left\langle {E\bf }^2\right\rangle/8\pi$ is the thermal component of the electric energy density in vacuum.
The \emph{dispersive} component
of the force (\ref{HenkelForce})
can be explicitly worked out by introducing the
Fourier transform  $g_{ik}[\omega;{\bf K},z_a,z_b]$ of the
Green function $G_{ik}[\omega;{\bf r}_a,{\bf r}_b]$
where ${\bf K}$ is the component of the
electromagnetic wavevector  parallel to  the interface.
By explicitly expressing the function $g_{ik}$  in terms of the
transmitted Fresnel coefficients \cite{Sipe}
and using the procedure described in \cite{henkelbis} we find,
after some  lengthy algebra,  the important result
%
\begin{gather}
F_{\text{th}}^{\text{neq}}(T,0,z)=-\frac{2\sqrt{2}\hbar \alpha _{0}}{\pi
\;c^{4}}\int_{0}^{\infty }\text{d}\omega \frac{\omega ^{4}}{e^{\hbar \omega
/k_{B}T}-1} \notag\\
\int_{1}^{\infty }\text{d}q\;qe^{-2z\frac{\omega }{c}\sqrt{q^{2}-1}}\;\sqrt{%
q^{2}-1}\sqrt{|\varepsilon (\omega )-q^{2}|+(\varepsilon ^{\prime }(\omega
)-q^{2})} \notag\\
\left( \frac{1}{\left| \sqrt{\varepsilon (\omega )-q^{2}}+\sqrt{1-q^{2}}%
\right| ^{2}}+\;\frac{\left( 2q^{2}-1\right) \left( q^{2}+|\varepsilon
(\omega )-q^{2}|\right) }{\left| \sqrt{\varepsilon (\omega )-q^{2}}%
+\varepsilon (\omega )\sqrt{1-q^{2}}\right| ^{2}}\right) .
\label{explicit}
\end{gather}
%
where we have introduced the dimensionless variable $q=Kc/\omega$.
Eq. (\ref{explicit})  provides the thermal force generated by
the substrate in the absence of  the environment radiation.
 In order to discuss the more general case
$T_S\neq T_E\neq 0$, we make use of the additivity property
of the  thermal force which can be written,
in general, as the sum of two
 contributions:
$F^{\textrm{neq}}_{\textrm{th}}(T_S,T_E,z)=F_{\textrm{th}}^{\textrm{neq}}
(T_S,0,z)+F_{\textrm{th}}^{\textrm{neq}}(0,T_E,z)$, produced, respectively, by the radiation of the substrate and of the environment.
Their additivity  can be checked  by evaluating
separately the two contributions and verifying that their sum,
at thermal equilibrium, reproduces the Lifshitz force $F^{\textrm{eq}}_{\textrm{th}}$ \cite{antezzafuture}.  
The full surface-atom force out of equilibrium can  be
finally written in the convenient form
\begin{equation}
F^{\textrm{neq}}(T_S,T_E,z)= F^{\textrm{eq}}
(T_E,z)+ F_{\textrm{th}}^{\textrm{neq}}(T_S,0,z)-
F_{\textrm{th}}^{\textrm{neq}}(T_E,0,z)\,
\label{fullnoneq}
\end{equation}
where the equilibrium force $F^{eq}(T,z)$  is given by (\ref{Feq}) while $F_{\textrm{th}}^{\textrm{neq}}(T,0,z)$ is
defined by eq.(\ref{explicit}). 
\begin{figure}[ptb]
\begin{center}
\includegraphics[width=0.44\textwidth]{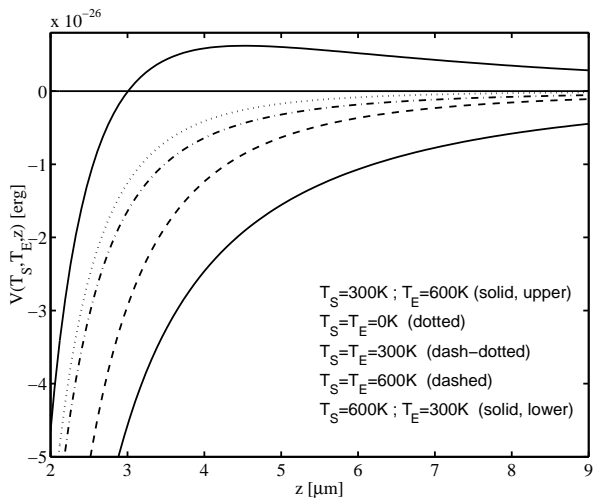}
\caption{\footnotesize Surface-atom potential
energy $V^{\textrm{neq}}(z)=-\int_z^{+\infty} dz' F^{\textrm{neq}}(z')$
calculated from eq.(\ref{fullnoneq}), for different
thermal configurations.}
\label{fig:1}
\end{center}
\end{figure}
\begin{figure}[ptb]
\begin{center}
\includegraphics[width=0.44\textwidth]{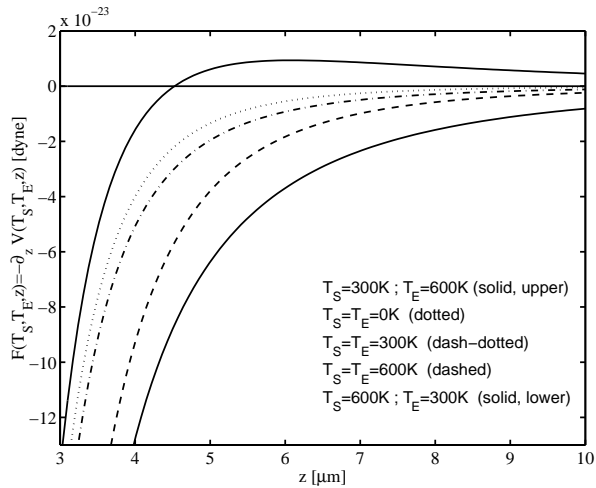}
\caption{\footnotesize Surface-atom force $F^{\textrm{neq}}(z)$
calculated from eq.(\ref{fullnoneq}), for different
thermal configurations. }
\label{fig:2}
\end{center}
\end{figure}
\begin{figure}[ptb]
\begin{center}
\includegraphics[width=0.44\textwidth]{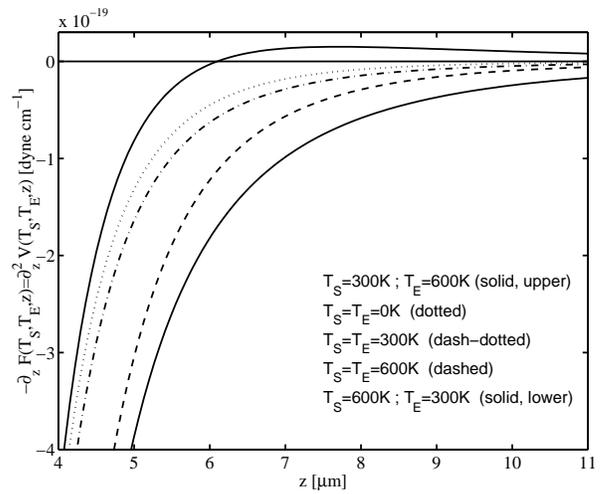}
\caption{\footnotesize Surface-atom gradient of the force
$\partial F^{\textrm{neq}}(z)/\partial z$ calculated
from eq.(\ref{fullnoneq}), for different thermal configurations. }
\label{fig:3}
\end{center}
\end{figure}

In figs. \ref{fig:1}, \ref{fig:2} and \ref{fig:3}
we show the explicit results for
the potential energy, the force and the gradient of the
force obtained starting from eq.(\ref{fullnoneq}) as a
function of the distance from the surface for different choices of $T_S$ and $T_E$.
Calculations have been done for a sapphire substrate and for rubidium atoms. For $F^{\textrm{eq}}(T,z)$ we have used the predictions of \cite{articolo1}. The figure clearly shows that the thermal effects out of equilibrium are sizable (solid lines), thereby providing promising perspectives for future measurements of the surface atom force at large distances. In particular in order to increase the attractive nature of the force it is much more convenient to heat the substrate by keeping the environment at room temperature (lower solid line) rather than heating the whole system (dashed line). When $T_S<T_E$ (upper solid line) the force exhibits a characteristic
change of sign reflecting a repulsive nature  at large distances (see also discussion below). At short distances the thermal correction to the force becomes smaller and smaller and is determined by the temperature of the substrate. 
 We have reported the  results for the potential,
for the force and for the gradient of the force because the
corresponding predictions can be of interest for different
types of experiments with ultra cold gases.
Experiments based on the study of the center of mass oscillation of a trapped gas 
are sensitive
to the gradient of the force \cite{articolo1}. The corresponding frequency shifts produced by the surface-atom interaction have been recently measured \cite{eric05} in conditions of thermal equilibrium in agreement with the predictions of theory \cite{articolo1}. Conversely  experiments based
on Bloch oscillations  are
sensitive to the force itself \cite{firenze,Iacopo}. Finally one can also
think at interference experiments with
Bose-Einstein condensates in a double well potential.
For large separations between the wells the position of
the corresponding interference fringes  are
sensitive to the potential \cite{Ketterle2}.

In the last part of the letter we discuss the large $z$
behaviour of the out of equilibrium force.
In this limit only values $q\approx 1$ are
important in the evaluation of the integral (\ref{explicit}).
By making the substitution $q^{2}-1=t^{2},$ and  the $t\ll 1$
expansion we find  that the  force (\ref{explicit})
exhibits the non trivial asymptotic behaviour
\begin{gather}
{F}^{\textrm{neq}}_{\textrm{th}}(T,0,z)_{z\to \infty}
=-\frac{\hbar \alpha_0}{z^{3}\pi \;c}%
\int_{0}^{\infty }\text{d}\omega \frac{\omega }{e^{\hbar \omega /k_BT}-1}%
f\left( \omega \right) .
\label{LD}
\end{gather}
Notice that  the force exhibits a slower $1/z^3$
decay with respect to  the one holding at thermal
equilibrium where it  decays like $1/z^4$
(see eq. (\ref{neweqasymp})). In the above equation
we have introduced the function
\begin{equation}
f\left( \omega \right)=\left( |\varepsilon(\omega)-1| +
( \varepsilon^{\prime }(\omega)-1) \right)^\frac{1}{2}
\frac{ 2+|\varepsilon(\omega)-1|}{\sqrt2 |\varepsilon(\omega)-1|}
\label{f}
\end{equation}
which depends on the optical properties of the substrate.
For temperatures much smaller than the energy
$\hbar \omega_c/k_B$,   where $\omega_c$ is the lowest
characteristic frequency of the dielectric substrate,
we can replace $f(\omega)$ with its low frequency
limit $(\varepsilon_0+1)/\sqrt{\varepsilon_0-1}$.
The  force (\ref{fullnoneq}) felt
by the atom then approaches the asymptotic behaviour
\begin{equation}
F^{\textrm{neq}}(T_S,T_E,z)_{z \to \infty} =
-\frac{\pi}{6}\frac{\alpha _{0}k_B^2(T_S^2-T_E^2)}{z^{3}\;
c\hbar}\frac{\varepsilon _{0}+1}{\sqrt{\varepsilon _{0}-1}}
\label{LevLimit}
\end{equation}
holding  at low temperature and at distances larger than
$\lambda_T/\sqrt{\varepsilon_0-1}$
 where $\lambda_T$ is the thermal photon wave
length calculated at the relevant temperatures
 $T_S$ and $T_E$ \cite{static}. Eq.(\ref{LevLimit})
shows that, at large distances, the new force  is
attractive or repulsive depending on whether the
substrate temperature  is higher or smaller than the
environment one.
Furthermore it exhibits a stronger temperature dependence with respect to equilibrium and
contains explicitly the Planck constant. The new dependence of $F^{\textrm{neq}}(T,0,z)$
on temperature and distance can  be physically understood by noticing that the main contribution  to the $z$-th dependent part of the electric 
energy $U_E$ arises from the black-body radiation impinging on the surface in a small interval of angles, of order of $(\lambda_T/z)^2$, near the angle of total reflection. This radiation creates slowly damping evanescent waves in vacuum. As a result $F^{\textrm{neq}}(T,0,z)$ turns out to be, in accordance with 
eq.(\ref{LevLimit}), of order of $-(\lambda_T^2/z^3)U_{BB}$, where $U_{BB}\propto T^4$ is the energy density of the black-body radiation.

Equation (\ref{LevLimit}) holds for a dielectric substrate
where $\varepsilon_0$ is finite.
For a metal, if one uses the Drude model, one has
 $\varepsilon^{\prime \prime }(\omega )
=4\pi \sigma /\omega$ with the real part $\varepsilon^{\prime}(\omega)$
remaining finite as $\omega\to 0$ so that
 one finds $f\left(\omega \right) \rightarrow \sqrt {\varepsilon
^{\prime \prime }(\omega )/2}=\sqrt{2\pi \sigma/\omega }$.
At low temperatures eq.(\ref{LD})
then gives rise to a different  temperature dependence 
\begin{equation}
F^{\textrm{neq}}(T_S,T_E,z)_{z\to \infty}=
- \frac{\alpha_0\zeta(3/2)\sqrt{\sigma}k_B^{3/2}(T_S^{3/2}-
T_E^{3/2})}{z^{3}\;c\sqrt{2\hbar}}
\label{LevLimMetal}
\end{equation}
where  $\zeta(3/2)\sim 2.61$ is the usual Riemann function.

In conclusion in this Letter we have calculated the surface-atom force out of thermal equilibrium and pointed out the occurrence of a new asymptotic behaviour at large distances. Our predictions could be tested in experiments with ultracold atomic gases trapped close to the surface of a substrate.

We are grateful to E. Cornell, J. Obrecht,
J. McGuirk and D.M. Harber for many useful comments.
It is also a pleasure to thank I. Carusotto,
C. Henkel and S. Reynaud for insightful discussions.
Financial support from MURST is acknowledged.



\begin{thebibliography}{99}
\bibitem{static} The static limit, which is obtained by replacing
$\varepsilon(\omega)$ with $\varepsilon_0$,
can be also used to simplify the evaluation of eq. (\ref{explicit}).
In the case of a sapphire substrate at ordinary temperatures ($\varepsilon_0=9.4$)
this approximation is very accurate at all distances.
\end{thebibliography}
\end{document}